\documentclass[a4paper,fleqn,usenatbib]{mnras}

\usepackage[T1]{fontenc}
\usepackage{ae,aecompl}

\usepackage{graphicx}	
\usepackage{amsmath}	
\usepackage{amssymb}	

\def\fbbh{f_{\bullet \bullet}} \def\msun{M_\odot}
\def\kms{\rm km\,s^{-1}}

\title[Host galaxy properties of gravitational wave sources]{Host
galaxy properties of mergers of stellar binary black holes and their
implications for advanced LIGO gravitational wave sources}
\author[L. Cao et al.]{ Liang Cao,$^{1,2}$ Youjun
Lu,$^{1,2}$\thanks{E-mail: luyj@nao.cas.cn} and Yuetong Zhao$^{1,2}$
\\
$^{1}$National Astronomical Observatories, Chinese Academy of
Sciences, 20A Datun Road, Beijing 100012, China\\ $^{2}$School of
Astronomy and Space Sciences, University of Chinese Academy of
Sciences, 19A Yuquan Road, Beijing 100049, China }
\date{Accepted 2017 November 24; Received 2017 November 23; in original form 2017 October 9}

\pubyear{2017}

\begin{document}

\label{firstpage}

\pagerange{\pageref{firstpage}--\pageref{lastpage}}

\maketitle

\begin{abstract}
Understanding the host galaxy properties of stellar binary black hole
(SBBH) mergers is important for revealing the origin of the SBBH
gravitational-wave sources detected by advanced LIGO and helpful for
identifying their electromagnetic counterparts. Here we
present a comprehensive analysis of the host galaxy properties of
SBBHs by implementing semi-analytical recipes for SBBH formation and
merger into cosmological galaxy formation model.  If the time delay
between SBBH formation and merger ranges from $\la$\,Gyr to the Hubble
time, SBBH mergers at redshift $z\la0.3$ occur preferentially in big
galaxies with stellar mass $M_*\ga2\times10^{10}\msun$ and
metallicities $Z$ peaking at $\sim0.6Z_\odot$. However, the host
galaxy stellar mass distribution of heavy SBBH mergers
($M_{\bullet\bullet}\ga50\msun$) is bimodal with one peak at
$\sim10^9\msun$ and the other peak at $\sim2\times10^{10}\msun$.  The
contribution fraction from host galaxies with $Z\la0.2Z_\odot$ to
heavy mergers is much larger than that to less heavy mergers.  If
SBBHs were formed in the early universe (e.g., $z>6$), their mergers
detected at $z\la0.3$ occur preferentially in even more massive
galaxies with $M_*>3\times10^{10}\msun$ and in galaxies with
metallicities mostly $\ga0.2Z_\odot$ and peaking at $Z\sim0.6Z_\odot$,
due to later cosmic assembly and enrichment of their host galaxies.
SBBH mergers at $z\la0.3$ mainly occur in spiral galaxies, but the
fraction of SBBH mergers occur in elliptical galaxies can be
significant if those SBBHs were formed in the early universe; and
about two thirds of those mergers occur in the central
galaxies of dark matter halos. We also present results on the host
galaxy properties of SBBH mergers at higher redshift.
\end{abstract}




\section{Introduction}
\label{sec:intro}

Gravitational wave (GW) events from mergers of stellar binary black
holes (SBBHs) are now expected to be regularly detected by the
advanced Laser Interferometer Gravitational Observatory (aLIGO),
examples include GW150914 \citep{2016PhRvL.116f1102A}, GW151226
\citep{16Abbott1226}, GW170104 \citep{GW170104}, and GW170814
\citep{GW170814}. These detections not only confirm the existence of
GWs and demonstrate the existence of SBBHs in the universe, but also
offer a great tool to study the astrophysical origin of SBBHs and the
abundant stellar and dynamical physics involved in, which are still
not well understood, yet.

A number of mechanisms have been proposed to produce SBBH GW sources
\citep[e.g.,][]{2002ApJ...572..407B,2015PhRvL.115e1101R,
2016PhRvD..93h4029R, 2016MNRAS.459.3432M,Wang16, 2016PhRvL.117f1101S,
Wanghuang16, McKernan17}, among which the evolution of massive binary
stars in galactic fields in isolation is possibly the leading
mechanism \citep[e.g.,][]{2016MNRAS.458.2634M, 2016Natur.534..512B}.
According to comprehensive population synthesis modeling of the
formation of SBBHs from binary zero-age main-sequence (ZAMS) stars
\citep[][for other SBBH population synthesis model, see also
\citealt{1996A&A...309..179P, Spera15, 2017MNRAS.470.4739S,
2000MNRAS.315..543H, 2002MNRAS.329..897H}]{2016Natur.534..512B},
GW150914-like sources, i.e., mergers of heavy SBBHs, are required to
be formed in metal poor environments (with metallicity $Z\la 0.1
Z_\odot$) at either an early time of the universe (with a
corresponding redshift of $z \ga 3$) or very recently ($z\sim 0.2$)
\citep[see also][]{2016PhRvL.116f1102A, 2016MNRAS.460L..74H,
Lamberts16, OShaughnessy17, Elbert17, Schneider17}. If this is true,
then the host galaxies of those GW150914-like sources at the GW
detection time may be either massive or small
\citep[e.g.,][]{OShaughnessy10,Lamberts16, OShaughnessy17, Elbert17,
Schneider17}.  It is anticipated that the properties of the host
galaxies of GW sources, if identified by future observations, can be
used to reveal the formation mechanism for SBBHs and constrain the
physics involved in the SBBH formation processes.

To identify the origin of GW sources, one of the crucial ways is to
find their electro-magnetic (EM) counterparts, if any, since the
localization of these sources, obtained from GW signals only, is poor
(typically covering a sky area of hundreds of square degrees with more
than ten thousands of galaxies in it) \citep{Abbott16LRR}. Great
efforts have been put into searching for EM counterparts of GW sources
via broadband campaign (e.g., GW150914). After the submission of this
paper, GW signals emitted from a binary neutron star merger (GW170817)
have been detected by aLIGO and Virgo with greatly improved
localization (\citet{2017PhRvL.119p1101A}), and its EM counterpart, a
short gamma ray burst (GRB170817A) and kilonova, is found by
subsequent multiband observations
(\citet[e.g.,][]{2017ApJ...848L..13A}. However there still seems
little expectation of the detection of EM counterparts for SBBH
mergers \citep{Abbott16d}. If the host galaxy properties of GW sources
can be known, the search for EM counterparts would be greatly
narrowed.  Therefore, it is of great importance to figure out where
and when the GW sources were formed and what kind of galaxies they are
hosted in at the GW detection time by the means of searching for the
EM counterparts of SBBH GW sources.

Some attempts have been made on investigating the host galaxies of
SBBH GW sources \citep[e.g.,][]{Lamberts16, OShaughnessy17, Elbert17,
Schneider17} since the discovery of GW150914. By combining
observational estimates of galaxy properties across cosmic time and
considering the effects of galaxy mergers, \citet{Lamberts16} found
that the GW150914-like sources may be dominated by binaries in massive
galaxies formed at early time (with $z\simeq 2$) if their progenitors
have metallicity $Z \geq 0.1 Z_\odot$ while come from binaries in
dwarf galaxies formed at later time (with $z\simeq 0.5$) if $Z <
0.1Z_\odot$, which appears somewhat different from
\citet{Belczynski16}. Using detailed cosmological simulations of Milky
way-like halos, \citet{Schneider17} found that GW150914-like sources
may be formed in low-metallicity dwarf galaxies at high redshift
($2.4\leq z\leq 4.2$) but be mostly hosted by star forming galaxies
with mass $>10^{10}M_\odot$. \citet{Elbert17} also investigated the
host galaxy properties of GW sources by using the observed properties
of galaxies and found that SBBH mergers would be mostly localized in
dwarf galaxies if the merger timescale is short but in massive
galaxies otherwise. Difference among the results output from the above
different approaches may be due to the use of different input models
for the BH-BH properties, the limitations (or uncertainties) in using
scaling relations and its extrapolations, or lack of information on
SBBHs formed in each galaxy \citep[see discussions in][]{Schneider17}.
Although \citet{Schneider17} did consider the SBBH formation history
in individual galaxies by using cosmological simulations but it is
limited to a small volume (size of $4$\,cMpc) and  halos with mass up
to Milky way-size.  \citet{Mapelli17} recently considered the SBBH
formation by using the Illustris simulation of galaxy formation in a
large box ($106.5$\,Mpc), but focused on the the merger rate density
evolution and the effect of metallicity.

In this paper, we investigate the properties of the host galaxies of
the SBBH GW sources by implementing simple SBBH formation recipes into
a cosmological galaxy formation model. In this approach, the
cosmological N-body  Millennium-II simulation in a box of side
$137$\,Mpc \citep{2009MNRAS.398.1150B}, semi-analytical galaxy
formation recipes \citep{Guo11}, and simple recipes for SBBH formation
following the star formation in each galaxy are combined together.
Therefore, the SBBH formation histories in a large number of
individual galaxies with masses from $10^7 M_\odot$ to several times
of $10^{11} M_\odot$ are also resolved. In section~\ref{sec:form}, we
describe the model for SBBH formation and SBBH mergers by utilizing
the cosmological galaxy formation model results presented in
\citet{Guo11} and simple recipes for SBBH formation from the evolution
of binary stars and SBBH mergers since then in each individual model
galaxies.  We generate mock catalogs for SBBHs, SBBH mergers, and
their host galaxies in section~\ref{sec:mock}. According to the mock
samples, statistics on the properties of the host galaxies of SBBH GW
sources are obtained and presented in section~\ref{sec:cal}.
Conclusions and discussions are given in section~\ref{sec:cons}.

\section{Binary Black Holes formation and merger scenario}
\label{sec:form}

In general,  the birth rate $R_{\rm birth}$ of single BHs with mass
$m_{\bullet}$ per unit volume per unit time at the cosmic time $t$ can
be estimated by \citep[see also][]{2016PhRvL.116m1102A,
2016MNRAS.461.3877D},
\begin{eqnarray}
R_{\rm birth}(m_{\bullet}, t) = \int\int \dot{\psi}[Z;
t-\tau(m_\star)] \phi(m_\star) \times \nonumber \\
\delta(m_\star-g^{-1}_{\bullet}(m_\bullet,Z)) dm_\star dZ.
\label{eq:bbh_birthrate}
\end{eqnarray}
Here $\dot{\psi}(Z; t)$ is the star formation rate with  metallicity
$Z$ per unit volume per unit time at the cosmic time $t$,
$\tau(m_\star)$ is the lifetime of a star with mass $m_\star$,
$\phi(m_\star) \propto m_\star^{-\alpha}$ is the initial mass function
(IMF) and the Chabrier IMF \citep{Chabrier03} is adopted
below,\footnote{Note here that the dependence of the black hole
formation on the assumed IMF is weak \citep{Elbert17}.} $\delta$ is
the Dirac-$\delta$ function, $m_{\bullet} = g_{\bullet}(m_\star,Z)$ is
a function that describes the relation between the mass of a stellar
remnant BH and the mass of its progenitor star and the latest version
given in \citet{Spera15} is adopted in this paper. Note that the
results on the relation between remnant BH mass and progenitor mass
obtained by \citet{1995ApJS..101..181W} and
\citet{2012ApJ...749...91F} are also frequently adopted in the
literature, but alternatively adopting those results does not lead to
significant effects on our conclusions \citep[see also discussions
in][]{Elbert17}.  Since the evolution time of BH progenitors are just
a few times $10^6$\,yr, which is negligible compared with the
evolution time of galaxies, therefore, we ignore $\tau(m_\star)$ in
the following calculations.

Considering SBBHs formed from isolated massive binary stars in
galactic fields, only a fraction ($f_{\rm b,*}$) of stars are in
binaries, a faction ($\fbbh$) of binary that can evolve to SBBHs, a
fraction ($f_q$) of SBBHs that have large mass ratio
$q=m_{\bullet,2}/m_{\bullet,1}$ with $m_{\bullet,1}$ and
$m_{\bullet,2}$ the masses of the primary component and the secondary
one, respectively, and a fraction ($f_{\rm mrg}$) of  them that can
merger within the Hubble time. An SBBH may merger after a time period
of $t_{\rm d}$ since its formation due to its orbit decay by GW
radiation. The GW event rate is then given by the convolution of the
birthrate $R_{\rm birth}(m_{\bullet,1},q;t)$ with the delay time
distribution $P(t_{\rm d})$ (see a similar approach by
\citet{2016MNRAS.461.3877D}), i.e.,
\begin{equation}
R_{\rm GW}(m_{\bullet,1},q;t) = f_{\rm eff} \int_{t_{\rm l}}^{t_{\rm
u}} R_{\rm birth}(m_{\bullet,1},t-t_{\rm d})  P_q(q) P_t(t_{\rm
d})dt_{\rm d},
\label{eq:bbh_birthrate2}
\end{equation}
and
\begin{equation}
f_{\rm eff} = f_{\rm b,*}  \times f_{q} \times \fbbh \times f_{\rm
mrg}.
\end{equation}
Here the distribution of mass ratio $P_q(q)$ is  assumed to be
independent of the BH mass and is normalized as $\int_{q_{\rm min}}^1
P_q(q) dq = 1$, $f_{\rm eff}$ is the effective factor to form GW
sources from binary stars, $t_{\rm l} =t_{\rm d,min}$ is the minimum
time delay, $t_{\rm u}= \int^{\infty}_{z(t)} \left|
\frac{dt}{dz}\right| dz$ is the longest possible time delay for an
SBBH merger occurred at the cosmic time $t$ and correspondingly
redshift $z$.

As seen from equations~(\ref{eq:bbh_birthrate}) and
(\ref{eq:bbh_birthrate2}), the event rate of SBBH GW sources depends
on not only the (binary) star formation rate, but also the mass ratio
distribution $P_q(q)$ and the distribution of the delay time
$P_t(t_{\rm d})$ between the binary formation and the final merger.
Distributions $P_q$ and $P_t$ depend on detailed physics involved in
the evolution processes of massive binary stars towards SBBHs and are
expected to not directly depend on their environment at large scales,
e.g., host galaxies (see sections~\ref{sec:pt} and \ref{sec:pq}). The
star formation rate $\dot{\psi}(Z;t)$ is the cosmic mean star
formation rate averaged over all galaxies, which can be obtained from
observations (e.g., as that adopted in \citealt{2016MNRAS.461.3877D}).
Considering that the star formation histories of different GW host
galaxies can be significantly different, therefore,
equation~(\ref{eq:bbh_birthrate}) can be replaced by
\begin{eqnarray}
R_{\rm birth}(m_{\bullet}, t) & = & \frac{1}{\Delta V \delta t}
\sum_{i=1}^{N} \mathcal{R}_{\rm birth,i}(m_{\bullet},t), \\
\mathcal{R}_{\rm birth,i}(m_{\bullet},t) & = & \int \int \sum_j
\Psi_{ij} [Z_{ij}; t-\tau(m_\star)] \phi(m_\star) \times \nonumber \\
 & & \delta(m_\star-g^{-1}_{\bullet}(m_\bullet,Z_{ij})) dm_\star
dZ_{ij}.
\label{eq:bbh_birthrate3}
\end{eqnarray}
Here $\Delta V$ is the comoving volume of the observable universe or a
comoving volume that is considered (e.g., a simulation box) at the
cosmic time from $t$ to $t+\delta t$ , $\mathcal{R}_{\rm birth,i}$ is
the total number of single BHs with mass $m_{\bullet} \rightarrow
m_{\bullet} + d m_{\bullet}$ over a time period of $t \rightarrow t
+\delta t$ in a galaxy $i$ or its progenitors ($t$ is a time earlier
than the detection time of the galaxy $i$), $i=1, ..., N$ indicate all
galaxies in the volume $\Delta V$, $\Psi_{ij}$ represents the total
mass of stars formed in galaxy $j$ in the time period $t \rightarrow
t+ \delta t$, one of the progenitor galaxies of galaxy $i$, $Z_{ij}$
is the metallicity of the star forming gas in the progenitor galaxy
$j$, and the summation is for all progenitor galaxies of the galaxy
$i$. The GW event rate of  a single galaxy $i$ at the cosmic time $t$
is given by
\begin{equation}
\mathcal{R}_{\rm GW,i}(m_{\bullet,1},q;t) \propto \int_{t_{\rm
l}}^{t_{\rm u}} R_{\rm birth,i}(m_{\bullet,1},t-t_{\rm d})  P_q(q)
P_t(t_{\rm d})dt_{\rm d},
\label{eq:GWratei}
\end{equation}
and the mean GW event rate per unit volume per unit time is given by
the summation of the contribution from all galaxies
\begin{equation}
R_{\rm GW}(m_{\bullet,1},q;t) = f_{\rm eff} \sum_{i=1}^N
\mathcal{R}_{\rm GW,i}(m_{\bullet,1},q;t).
\label{eq:GWratesumi}
\end{equation}

\subsection{$\Psi_{ij}(Z_{ij};t)$ for individual mock galaxies from a
cosmological galaxy formation model}

In order to study the host galaxy properties of those GW events, we
use the catalogs of mock galaxies obtained from semi-analytical galaxy
modeling by \citet{Guo11}, in which the assembly and star formation
histories of each individual mock galaxy and its progenitors are
given. These catalogs are obtained from the halo/subhalo merger trees
of the Millennium II simulation \citep{2009MNRAS.398.1150B} by
implementing parameterized semi-analytical modeling of the galaxy
formation recipes of a variety of physics processes,e.g. a
mass-dependent model for supernova feedback, more realistic treatments
of gaseous and stellar disk growth, a updated reionization model. The
free parameters of these models are determined by using the observed
abundance, structure and clustering of low redshift galaxies as a
function of stellar mass \citep{Guo11}.

The outcome of the catalog is stored in a box with comoving size of
$137$\,Mpc for 68 snapshots with redshift from 127.0 to 0.  For each
individual galaxy in each snapshot, its properties, such as position,
stellar mass ($M_*$,$Z_*$), star formation rate ($\Psi_{ij}/\delta
t$), star forming gas metallicity ($Z_{ij}$), and its identity ID
(described by $i$, $j$, $t$) in the merger tree, are all recorded.
Therefore, how many SBBH mergers or GW events happened in the time
period $t\rightarrow t +\delta t$ in a galaxy can be obtained from the
amount of stars formed at the time $t-t_{\rm d}$ in the progenitor(s)
of that galaxy according to equation~(\ref{eq:bbh_birthrate3}).

\subsection{$P_t(t_{\rm d})$}
\label{sec:pt}

A significant fraction of SBBHs that merge at low redshift may have
actually formed in the early universe, therefore GW events can be used
to investigate the formation of BHs and SBBHs at a much earlier time
(see eq.~\ref{eq:bbh_birthrate2}), which are otherwise invisible to
current electromagnetic observations. The dependence of SBBH GW events
on the the star formation history is significantly affected by the
accuracy of the estimate of $P_t(t_{\rm d})$, which appears not well
understood because of various uncertainties in the binary evolution
model.

In most cases, $t_{\rm d}$ can be comparable to the age of the
Universe, and it generally depends on the initial orbital
configuration of binaries.  Theoretical studies have shown that $P_t
(t_{\rm d}) \propto t_{\rm d}^{-1}$
\citep[e.g.,][]{2008ApJ...672..479O, OShaughnessy10,
2016Natur.534..512B, Lamberts16}. The minimum time delay is on the
order of $\la$\,Gyr but with some uncertainties. In this paper, we
adopt two values for the minimum time delay $t_{\rm d, min}$, i.e.,
$50$\,Myr and $2$\,Gyr.  We denote the first one as the ``reference
model'' and the second one as the ``large time delay model''.  We also
adopt two extreme models for $P(t_{\rm d})$, one is the prompt model
with $P_t(t_{\rm d}) \propto \delta(t_{\rm d})$, i.e., SBBHs merge
right after their formation; the other one is a model for early
formation of SBBHs at redshift $z>6$ (denoted as the ``early formation
model''), for which $P_t(t_{\rm d}) \propto 1/t_{\rm d}$, $t_{\rm
d,min}= \max[t(z) - t(z=6), 50\,{\rm Myr}]$, and $t(z) =
\int^{\infty}_z \left| \frac{dt}{dz'} \right| dz'$ is the cosmic age
at redshift $z$.  The prompt model introduced in this paper is only
for the comparison with other models.  Assuming the prompt model, the
GW event rate should be roughly determined by the number of stars
formed at the GW detection time. In other three models, the GW event
rate is determined by the number of all stars formed earlier than the
time that is about $t_{\rm d,min}$ before the GW detection time.

The set of the ``early SBBH formation model'' is based on that the
detected heavy SBBHs, i.e., GW150914, GW170104, and GW 170814, may have to be
formed from metal poor massive binary stars at high redshift
\citep[$z\ga 3$; e.g.,][]{2016PhRvL.116m1102A, 2016Natur.534..512B, 2016MNRAS.460L..74H}.
It is also anticipated that a significant fraction of the future aLIGO
SBBH detections might be originated from those metal poor massive
binary stars formed at very high redshift, possibly Pop III stars, and
these SBBHs are probably hosted in highly biased systems.

\subsection{$P_q(q)$}
\label{sec:pq}

According to population synthesis models, SBBH mergers resulting from
evolution of massive binary stars typically have comparable mass
components and $P_q(q)$ peaks at $0.8-1.0$ \citep[e.g.,][]{deMink13,
Belczynski16, 2016Natur.534..512B}. The formation of low mass ratio
SBBHs is suppressed because low mass ratio binary stars, with
sufficiently small separations, tend to merge with each other before
they can evolve to SBBHs. Moreover, the mass ratio extends to lower
values with decreasing total mass of SBBHs. In principle, reliable
estimates of $P_q(q)$ may be obtained by intensive population
synthesis model calculations, which can be included in
equation~(\ref{eq:bbh_birthrate2}) to generate SBBHs. However, we note
here that the difference in $P_q(q)$ for SBBHs with different total
mass seems not very significant \citep[see][]{2016Natur.534..512B},
therefore, we assume $P_q(q) \propto q$ over the range from $0.5$ to
$1$, which seems to be consistent with population synthesis results
and compatible with the current aLIGO detections
\citep[e.g.,][]{2016PhRvL.116f1102A, 16Abbott1226, GW170104}.  We note
here that this setting does not affect our results on the host galaxy
properties since the mass ratio of SBBHs formed from massive binary
stars in galactic fields is independent of their environment at large
scale.

\begin{figure}
\begin{center}
\includegraphics[scale=0.60]{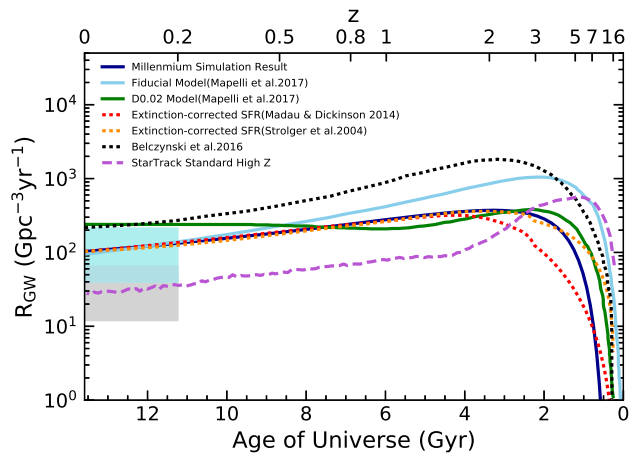}
\caption{
SBBH merger rate density as a function of the age of universe or
redshift $z$. The blue solid line represents the results obtained from
the Millennium simulation galaxy catalog assuming the reference model
for the time delay. The red and orange dotted lines represent our model results by using
the extinction-corrected specific star formation rate given by
\citet{MD14} and \citet{Strolger}, respectively, and the mean
metallicity redshift evolution in \citet{Belczynski16}.  The black
dotted line represents the result obtained by \citet{Belczynski16} by
using the extinction-corrected specific star formation rate in
\citet{MD14} and the binary population synthesis code StarTrack. The
green and light blue solid lines represent respectively the results
from the fiducial model and the D0.02 model in \citet{Mapelli17} by planting BBHs obtained from  the binary
population synthesis code SEVN into the Illustris simulations. The
purple dashed line represents the merger rate density obtained by
\citet{Dominik13}.
The cyan and grey shaded regions represent the constraints on the
merger rate density obtained from those detected GW sources by
assuming two different IMF, respectively \citep{GW170104}.
}
\label{fig:f1}
\end{center}
\end{figure}

With the above prescription, we can estimate the GW event rate of
SBBH mergers according to
equations~(\ref{eq:bbh_birthrate3})-(\ref{eq:GWratesumi}) by using the
catalog of mock galaxies with detailed assembly histories given by
\citet{Guo11}. For comparison, we can also estimate the GW event rate
of SBBH mergers by using observationally determined SFR and mean
metallicity evolution according to equation~(\ref{eq:bbh_birthrate2}).
To do this calculation, we adopt the extinction-corrected SFR obtained
by \citet[][see their Eq. 15]{MD14} and also the one obtained by
\citet[][See their  Eq. 5]{Strolger}, and we adopt the mean metallicity
redshift evolution as that in \citet[][]{Belczynski16}.

Note that in the above approach all the physics governing the evolution of
SBBHs are encoded in the three independent functions $f_{\rm eff}$,
$P_q(q)$ and $P_t(t_{\rm d})$, and any correlation between the mass
distribution of SBBHs and their merger times is ignored although there should be
such a correlation as a result of the selected evolutionary pathways followed by
massive BBHs
systems. However, there are still large uncertainties in the evolution models
of massive (binary) stars, especially, the large uncertainties in the understanding
of a number of physical processes, such as the common envelope evolution, the
kick from supernova explosion, and the mass transfer etc. \citep[e.g.,][]{Dominik13, Mapelli17, 2017arXiv170909197D}.
Different models may result in significantly different merger rate densities. Below we show that our simple
approach can gives similar merger rate density evolution compared against those obtained by using binary population
synthesis codes.

Current detections of GW\,150914, GW\,151226, and GW\,170104 have
already put a constraints on the merger rate of SBBHs in the local
universe as $40-213$ or $12-65\,{\rm Gpc^{-3}\,yr^{-1}}$ if assuming a
Salpeter IMF or uniform in log-distribution for the primary components
of progenitor binary stars  \citep[][see the cyan- and grey-shaded
regions in Fig.~\ref{fig:f1}]{GW170104}. We adopt that constraint on
the mean detection rate of $103\,{\rm Gpc^{-3}\,yr^{-1}}$ to calibrate
the merger rate density obtained from different models and thus
constrain the unknown value $f_{\rm eff}$. We find that $f_{\rm eff}$
should be $\sim8.0 \times 10^{-4}$, $\sim 1.1 \times 10^{-3}$, $\sim
9.9\times 10^{-2}$, and $\sim 1.2\times 10^{-3}$, for the reference
model, the large time delay model, the early SBBH formation model, and
the prompt model, respectively.  The value on $f_{\rm eff}$ for the
reference model is roughly consistent with that obtained in
\citet{Elbert17}. We note here that the distributions of the host
galaxy properties of SBBH mergers obtained in this paper for each
model are not affected by the actual value of $f_{\rm eff}$.

Figure~ \ref{fig:f1} shows the results on the merger rate density
distribution as a function of redshift by adopting the reference model for
$P_t(t_{\rm d})$. As seen from this figure, the merger rate density obtained
by using the mock galaxy catalog from \citet{Guo11} (blue line) is consistent
with that obtained by using the observationally determined SFR from
\citet{Strolger} (yellow dotted line) or from \citet{MD14} (red dotted line),
especially at redshift $z\la 2$. The substantial differences at higher redshift
are mainly due to the differences in the SFRs.

For comparison, we also plot a number of estimates for the merger rate density
evolution obtained in the literature by using the binary population synthesis
model in Figure~\ref{fig:f1}. These estimates include the one by
\citet{Belczynski16} using the observationally determined SFR and the binary
population synthesis code StarTrack (black dotted line), by \citet{Mapelli17} using
the SFR of individual galaxies resulting from the Illustris simulation and a new
version of binary evolution code BSE (green and cyan lines representing two
different models), by \citet{Dominik13} using the StarTrack code and mock
galaxies generated from Press-Schechter like formalism.
The results obtained by using the binary population synthesis models can be
substantially different depending on the settings of the model parameters
describing supernova explosion, common envelope evolution, mass transfer,
and the supernova kick, etc. It can be seen from Figure~\ref{fig:f1} that the shape
of the merger rate density evolution resulting from our simple prescription is
only slightly shallower than most of those obtained by using binary population synthesis
model at least at redshift $z\la2$ but steeper than the D0.02 model in \citet{Mapelli17}. At higher redshift $z>2-3$), our simple model results in comparable similar merger rate as
the D0.02 model in \citet{Mapelli17}, but it results in
relatively less SBBH mergers compared
with others shown in Figure~\ref{fig:f1} obtained from the population synthesis models, which may be partly due to
the ignoration of the correlation between the mass  distribution of SBBHs and their merger times and the different SFR and metallicity distribution adopted in
different models.

\section{Mock samples}
\label{sec:mock}

Using the mock galaxy catalogs and the assembly and star formation
history of each mock galaxy, we randomly assign SBBH merger GW events
according to the probability resulting from
equation~(\ref{eq:GWratei}) for individual galaxies across cosmic
time. We also impose that $\min[m_{\bullet,1}, m_{\bullet,2}] \geq
5\msun$, which is set by considering that all BHs measured dynamically
have masses $\ga 5\msun$ \citep[e.g.,][for the evidence of a mass gap
at $3-5\msun$]{Ozel10, Farr11}.  With these procedures, we can obtain
mock catalogs of SBBH GW events  at any given cosmic time $t$ (or
correspondingly redshift $z$) for given $P_t(t_{\rm d})$ and $P_q(q)$.
We generate a number of mock catalogs of SBBH GW events at a number of
redshifts (e.g., $z=0.3$, $1$, and $2$, respectively) that enables the
statistic studies on the properties of their host galaxies.  Each GW
event is characterized by the masses of the two components, i.e.,
$m_{\bullet,1}$ and $m_{\bullet,2}=q m_{\bullet,1}$, the merger time
$t(z)$, the SBBH formation time $t(z)-t_{\rm d}$, the position and
other properties (e.g., stellar mass $M_*$, metallicity $Z_*$, and
morphology) of its host galaxy, and the total mass ($M_{\rm halo}$) of
their dark matter halo at the merger time $t(z)$, and the properties
of its progenitor galaxy that the SBBHs formed in at the SBBH
formation time $t-t_{\rm d}$.

We note here that some of those mock SBBHs may have received
considerable kicks due to supernovae explosions and  moved away from
their birth places. \citet{Mandel16} estimated the kicks according to
available observations on the black hole X-ray binaries (BHXBs) and he
found that the kick velocity is not required to be $> 80\kms$,
although larger kicks are not completely ruled out, yet
\citep[see][]{Repetto12}. \citet{Mirabel16} also suggests that the
kick velocity should be small and insignificant according to the
kinematics of BHXBs.  If assuming a kick velocity of $80\,\kms$, most
of the SBBHs would not be able to climb out from the potential of
their host dark matter halos at the SBBH formation time, as the mass
of the host dark matter halos at the SBBH formation time is typically
larger than $10^{10}\msun$ with escape velocity $\ga 100\,{\rm
km\,s^{-1}}$.  Therefore, we ignore the possible offsets, if any, of
the SBBH mergers at the GW detection time from those galaxies they
formed in. However, we note that larger kick velocity is still
possible, with which some BBHs may locate at the outskirt of or even
climb out from the potential of their host galaxies and dark matter
halos they formed in, especially when those hosts are small as
\citet{Perna17} recently investigated in.

\section{Model Results}
\label{sec:cal}

\subsection{Stellar mass distributions of SBBH host galaxies}

\begin{figure}
\begin{center}
\includegraphics[scale=0.75]{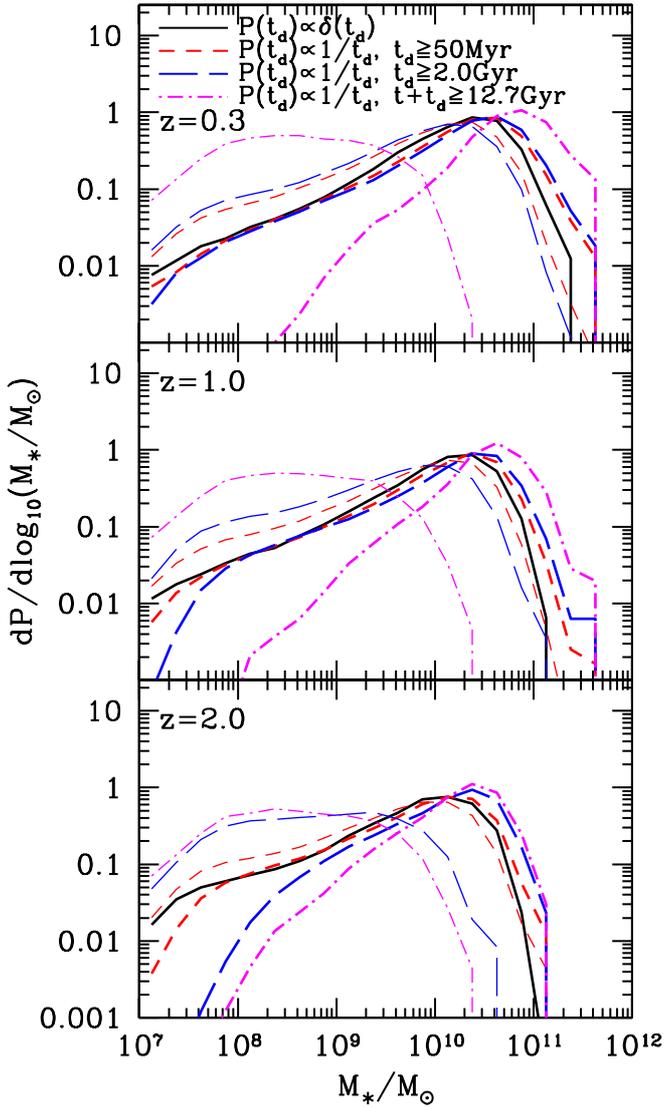}
\caption{Stellar mass distribution of the host galaxies of binary
black holes (SBBHs) at the SBBH formation time (thin lines) and merger
time/GW detection time (thick lines).  Top, middle and bottom panels
show those distributions for SBBHs if the GW signals due to their
mergers were detected at redshift $z=0.3$, $1$, and $2$, respectively.
In each panel, black solid line shows the results obtained form the
prompt model in which two SBBH components merge with each other right
after the SBBH formation. The red short-dashed, blue long-dashed, and
magenta dot-dashed lines show the results obtained form those time
delay models $P(t_{\rm d}) \propto 1/t_{\rm d}$ with $t_{\rm d} \ge
50$\,Myr, $\ge 2$\,Gyr, and $\ge 12.7{\rm \,Gyr}
- t(z)$, i.e., the reference model, the large time delay model, and
  the early SBBH formation model, respectively. Here
$t(z)=\int^{\infty}_z \left|\frac{dt}{dz}\right| dz$ and $12.7$\,Gyr
are the cosmic age at redshift $z$ and at $z=6$, respectively.
}
\label{fig:f2}
\end{center}
\end{figure}

We extract the statistical information on their host galaxies from the
mock catalogs for SBBH GW events and SBBHs. Figure~\ref{fig:f2} shows
the stellar mass distribution of the host galaxies of those SBBHs at
their merger time (i.e., the GW detection time $z=0.3$, $1.0$, and
$2.0$; thick lines) and formation time (thin lines), respectively. For
those GW events detected at low redshift (e.g., $z=0.3$, top panel),
the host stellar mass distribution at the SBBH formation time
resulting from the reference model (or the large time delay model)
peaks at $1.8\times10^{10} \msun$ (or $1.9\times10^{10}\msun $) with a
$5$ to $95$ percentile range of $1.7\times10^{8}$ to
$5.3\times10^{10}\msun$ (or $1.2\times10^{8}$ to $4.4\times10^{10}
\msun$)  (thin lines in the top panel), while at the SBBH merger time
it peaks at $3.3\times10^{10} \msun $ (or $3.3\times10^{10} \msun$)
with a $5$ to $95$ percentile range of $7.0\times10^{8}$ to
$9.8\times10^{10}\msun$ (or $7.6\times10^{8}$ to $1.1\times10^{11}
\msun$; thick line in the top panel).  The distribution of the SBBH
host galaxies at the GW detection time is shifted to significant
higher masses compared with that at the SBBH formation time simply
because of the growth of those host galaxies after the SBBH formation.
For those GW events detected at higher redshifts (e.g., $z=1, 2$),
similar trends are also seen, but both distributions at the SBBH
formation time and the GW detection time are substantially decrease at
the high mass end and the distribution peaks shift toward lower
masses.

\begin{figure}
\begin{center}
\includegraphics[scale=0.55]{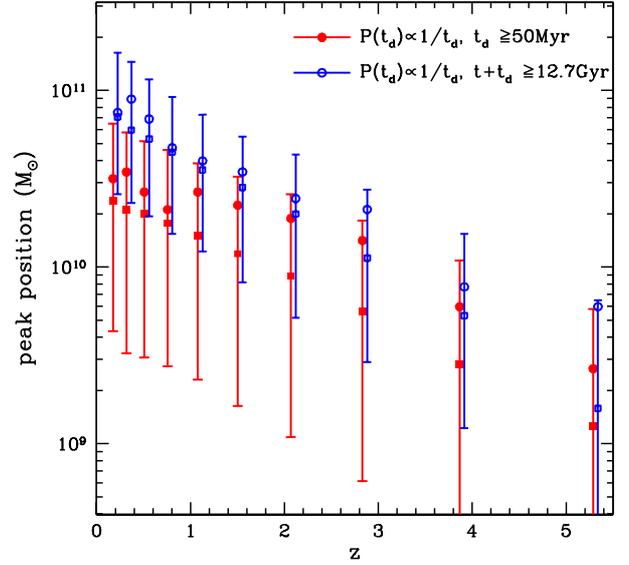}
\caption{Peak and median positions of the probability distribution of
the GW host galaxies as a function of merger redshift. The red solid
circles and squares show the peaks and the medians for the reference
model with $P(t_{\rm d}) \propto 1/t_{\rm d}$ and $t_{\rm d,min} \ge
50$\,Myr, and the blue open circles and squares show the peaks and the
medians for the early SBBH formation model with $P(t_{\rm d}) \propto
1/t_{\rm d}$ and $t_{\rm d,min} \ge 12.7{\rm \,Gyr}
- t(z)$, respectively. Vertical bars indicate the probability range of
  16\% percentile to 84\% percentile of host galaxies with mass from
low to high. Note that a small horizontal offset is added to each
point obtained from the early SBBH formation model, for clarity.}
\label{fig:f3}
\end{center}
\end{figure}

If the SBBHs were formed at early time, e.g., $z\ga 6$ (magenta
dot-dashed lines for the early SBBH formation model), their host
galaxies are  small at the SBBH formation time and the host stellar
mass distribution peaks at $2.8\times 10^{8} \msun$ with a $5$ to $95$
percentile range of $ 2.2\times 10^7$  to $3.5\times10^{9}\msun$;
while at the GW detection time, even at $z=2$ (magenta dot-dashed
lines in the bottom panel), their host galaxies grew significantly
since their formation and become massive, and the host stellar mass
distribution peaks at $2.2\times 10^{10} \msun$; the SBBH host
galaxies became even more massive if those SBBHs merged at $z=0.3$
(magenta dot-dashed line shown in the top panel), and the host stellar
mass distribution peaks at $7.5\times10^{10}\msun$ with a $5$ to $95$
percentile range of $7.5\times10^{9} $ to $2.1\times10^{11}\msun$. If
SBBHs merger right after their formation, i.e., the prompt model, the
host stellar mass distribution at the merger time is the same as that
at the formation time. Assuming this model, if the SBBH mergers were
detected at $z \sim 0.3$, their host galaxy mass distribution peaks at
$\sim 2.0\times10^{10}\msun$ and the peak position shifts slightly
towards lower masses with increasing detection redshift.

Figure~\ref{fig:f3} shows the peak and median of the host galaxy
stellar mass distributions of SBBH mergers at different GW detection
time.  The red filled circles and blue open circles represent those
resulting from the reference model and the early SBBH formation model,
i.e., $P(t_{\rm d}) \propto 1/t_{\rm d}$ with  $t_{\rm d} \ge
50$\,Myr, and $t_{\rm d} \ge 12.7{\rm \,Gyr} - t(z)$, respectively.
The bars associated with each point mark the $16$ and  $84$ percentile
of the distribution from low to high mass, respectively.  Apparently,
the peak and median masses of those distributions resulting from the
early SBBH formation model is substantially larger than those resulting
from the reference model, especially at low redshift (e.g., $z < 1.5$). Furthermore more,
the reference model results in a substantially wider range of the host
stellar mass at the GW detection time compared with the early SBBH
formation model.
The reason is as follows.
The host galaxies of those SBBH mergers in the early SBBH formation model
were mainly formed from high density peaks in the early universe, which grew up and collapsed into more massive objects in their subsequent evolution. The later the SBBH
mergers were detected, the more massive the SBBH host galaxies.
However, the host galaxies of those SBBH mergers in the reference model may be formed from both high density peaks in the early universe
and not so high density peaks at a time not long before the SBBH merger time.  Therefore, the mass distribution resulting from the reference model covers a wider range compared to that from the early SBBH
formation model, and the peak and median masses of the former one are also substantially smaller than those of the latter one, especially at low redshift, as shown in Figure~\ref{fig:f3}.
These results suggest that the mass distribution of the host galaxies of SBBH GW sources is
sensitive to the delay times and formation redshifts of these systems.

\subsection{Morphologies and types of SBBH host galaxies}

The host galaxies of GW events may have different types. Some SBBH GW
events may be hosted in elliptical galaxies and some others are hosted
in spiral galaxies. Some SBBH GW events may occur in central galaxies
or satellite galaxies in big halos, and some others may occur in
isolated galaxies in small dark matter halos or satellites in
sub-halos. Similar as that in \citet{Guo11}, elliptical and
spiral are discriminated by using the bulge mass to total stellar mass ratio. A
mock galaxy is labeled  as an elliptical if its bulge mass to total stellar mass ratio is larger than 0.8, and otherwise it is labeled as a spiral. If a mock galaxy is in the main subhalo of their host dark matter halo, it is labeled as the central galaxy of that halo; if it is in other subhalos, it is labeled as a satellite; and if it is the only galaxy in the halo, it is labeled as an isolated galaxy.

\begin{figure}
\includegraphics[scale=0.43]{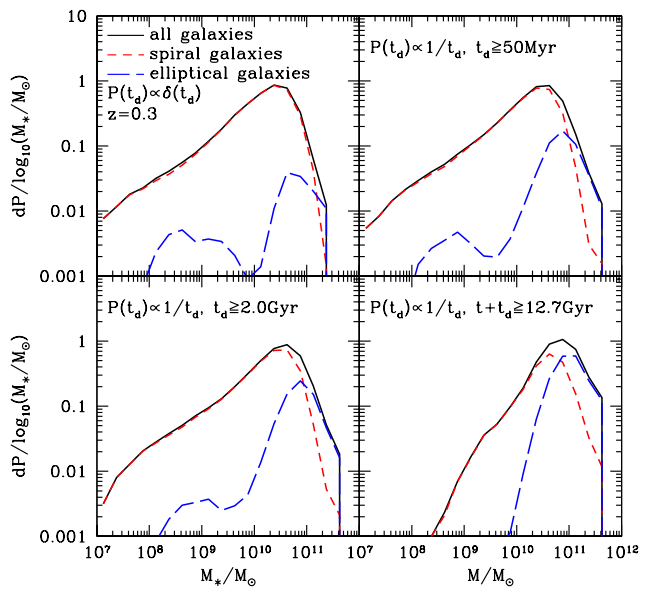}
\caption{Distributions of the host galaxies of those SBBH mergers as a
function of galaxy stellar mass for host galaxies with different
morphologies, including ellipticals and spirals.
}
\label{fig:f4}
\end{figure}

The fraction of those GW events that are hosted in ellipticals or
spirals depends on when and where those SBBHs were formed.
Figure~\ref{fig:f4} shows the probability distribution of those GW
host galaxies in different morphological types at the GW detection
time $z=0.3$ for different $P_t(t_{\rm d})$ models. As seen from this
Figure, if the time period between the SBBH formation and merger
distributed  in a broad range from $\sim 50$\,Myr to the Hubble time
(the reference model), the host galaxies of SBBH mergers with
$M_{\bullet\bullet} \ge 10\msun$ at  $z=0.3$ are mostly spiral
galaxies with the $5\%-95\%$ percentile stellar mass range of $\sim
5.3\times10^{8}-6.0\times 10^{10} \msun$, and only about 13\% are in
elliptical galaxies.  If those SBBHs were formed at early time (e.g.,
$z>6$, the early SBBH formation model) and detected through GW at
$z=0.3$, about 47\% of the GW events occur in elliptical galaxies with
mass $\ga 7.0\times10^{9}\msun $ and others are in spiral galaxies
typically with mass $\sim 4.2 \times 10^{9} - 9.4 \times 10^{10}
\msun$; if SBBHs merge shortly after their formation (i.e., the prompt
model), then $\sim 96\%$ of them should be hosted in spiral galaxies
with mass $\sim 5.3 \times10^{8} - 5.3 \times 10^{10} \msun$ and no
more than $4\%$ are in elliptical galaxies (top left panel). Apparently,
the longer the merger time, the larger the fraction of ellipticals in those
SBBH host galaxies. The reason is as follows. Spiral galaxies formed at
early time may merge into each other and form ellipticals in their later
evolution. The longer the evolution time, the larger the probability for
this morphology transformation. Therefore, the fraction of the
SBBHs mergers hosted in ellipticals grows with time if most
SBBHs were originally formed in spirals at early time.

\begin{figure}
\includegraphics[scale=0.43]{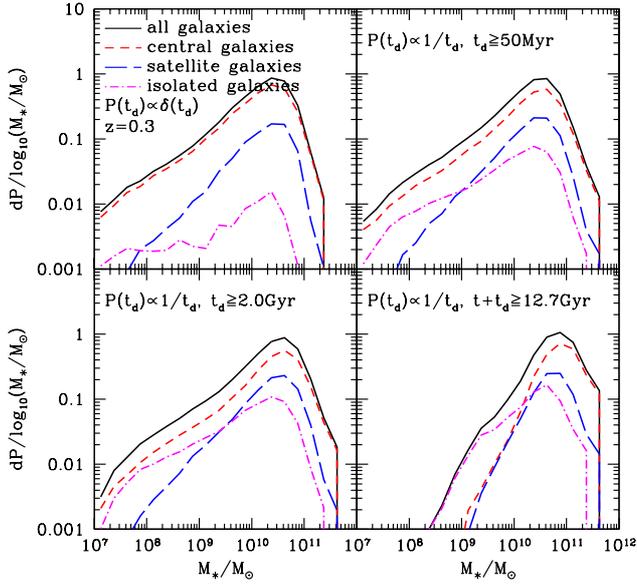}
\caption{Distributions of the host galaxies of those SBBH mergers as a
function of galaxy stellar mass for different types of host galaxies,
including central galaxies, satellites, and isolated galaxies.
}
\label{fig:f5}
\end{figure}

Figure~\ref{fig:f5} shows the probability distribution of those SBBH
GW events that are hosted in central galaxies, satellites, and
isolated galaxies, separately.  For four different
$P_t(t_{\rm d})$ models (shown in the top-left, top-right,
bottom-left, and bottom-right panels), the fraction that the SBBH GW
events are hosted in central galaxies (or satellites) are $80\%$ (or
$19\%$), $66\%$ (or $23\%$), $61\%$ (or $25\%$), and $62\%$ (or
$22\%$), respectively. As seen from this Figure, most of the SBBH
GW events are hosted in central galaxies. The reason is as follows.
The contribution from central galaxies to the star formation rate is
more significant compared with that from satellites and isolated galaxies.
Therefore, more SBBHs were formed in central galaxies. After the SBBH
formation, some of those central galaxies maintain as central galaxies of
their growing host halos, some of them merger with other central galaxies
or satellites to form new central galaxies, though a small fraction of those
central galaxies do sink into big halos to become satellites. At the SBBH
merger time, the host galaxies are still preferentially centrals, almost
independent of the delay time model.

\begin{figure}
\begin{center}:
\includegraphics[scale=0.43]{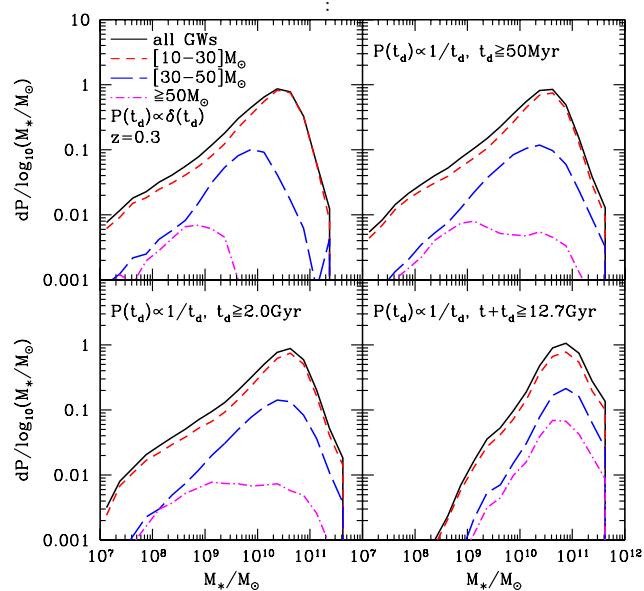}
\caption{Stellar mass distributions of the host galaxies for SBBH
mergers (at $z=0.3$) with different total masses.}
\label{fig:f6}
\end{center}
\end{figure}

\subsection{Host galaxies of SBBH mergers with different total masses}

SBBHs with different total masses [$M_{\bullet\bullet} = M_{\bullet,1}
+ M_{\bullet,2} = (1+q) M_{\bullet,1}$] may have different formation
histories.  Heavy SBBHs may be formed only from metal poor binaries
and thus in metal poor galaxies at early times, therefore, they should
be hosted in galaxies with different properties from those lighter
SBBHs at the SBBH merger time.

Figure~\ref{fig:f6} shows the stellar mass distributions of the host
galaxies of SBBH GW events with different $M_{\bullet\bullet}$ at z=0.3.  For
the prompt model, the host galaxies of those SBBH GW events with
$M_{\bullet\bullet} \geq 50\msun$, similar to GW150914, GW170104, and
GW170814, are small, with masses in the range from a few times
$10^7\msun $ to a few times $10^9\msun$ (the magenta line in top-left
panel of Fig.~\ref{fig:f6}); while those for SBBH GW events with
$M_{\bullet\bullet} \sim 30 -50\msun$ and $\sim 10 - 30\msun$ peak at
$6.7\times10^{9}\msun$ and $ \sim 2.4 \times 10^{10} \msun$ with a $5$
to $95$ percentile stellar mass range of $3.8\times10^{8}$ to
$2.7\times10^{10}\msun$ and $6.0\times10^{8}$ to
$6.0\times10^{10}\msun$, respectively (blue long-dashed and red
short-dashed lines).  If those SBBH GW events formed at early time
(e.g., $z\ga 6$ as the early SBBH formation model, bottom-right
panel), then the host stellar mass distribution of those SBBH GW
events with $M_{\bullet\bullet} \sim 30 -50\msun$ (or $\sim 10 -
30\msun$) peaks at $9.4\times10^{10}\msun$ (or
$7.5\times10^{10}\msun$) with a $5$ to $95$ percentile range of
$9.4\times10^{9}$ to $2.4\times10^{11}\msun$ (or $6.7\times10^{9}$ to
$2.1\times10^{11}\msun$);  while those heavy SBBHs with
$M_{\bullet\bullet}\geq 50\msun$, similar to GW150914 and GW170104,
are hosted in galaxies that have stellar masses around
$4.7\times10^{10}\msun$ with a $5$ to $95$ percentile range of
$6.0\times 10^9$ to  $2.4\times 10^{11}\msun$, similar to those for
the host galaxies of lighter SBBHs ($10 -50\msun$).

For other two $P_t(t_{\rm d})$ models with $t_{\rm d} \ge 50$\,Myr or
$\ge 2$\,Gyr, i.e., the reference model and the large time delay
model, the host galaxy stellar mass distribution of SBBHs with
different $M_{\bullet\bullet}$ range are in between those resulting
from the prompt model and the early SBBH formation model (with
formation time $z>6$). For those heavy SBBHs with $M_{\bullet\bullet}
\ge 50\msun$, the host stellar mass distribution are broad and bimodal
with one peak at $\sim 10^{9}\msun$ and the other peak at $\sim
2\times 10^{10}\msun$, which are corresponding to a population formed
recently in small galaxies and another population formed at early time
in small galaxies but merged into big galaxies later.

\subsection{Host galaxy metallicities}

Since the formation of BHs depends on the metallicity of their
progenitor stars, the formation of SBBHs is also dependent on the
metallicity of their progenitor binary stars and thus the metallicity
of the progenitor galaxies when they formed in. However, the
metallicity of the host galaxies at the GW detection time may be
significantly different from that of the progenitor galaxies at the
SBBH formation time.

\begin{figure}
\begin{center}
\includegraphics[scale=0.75]{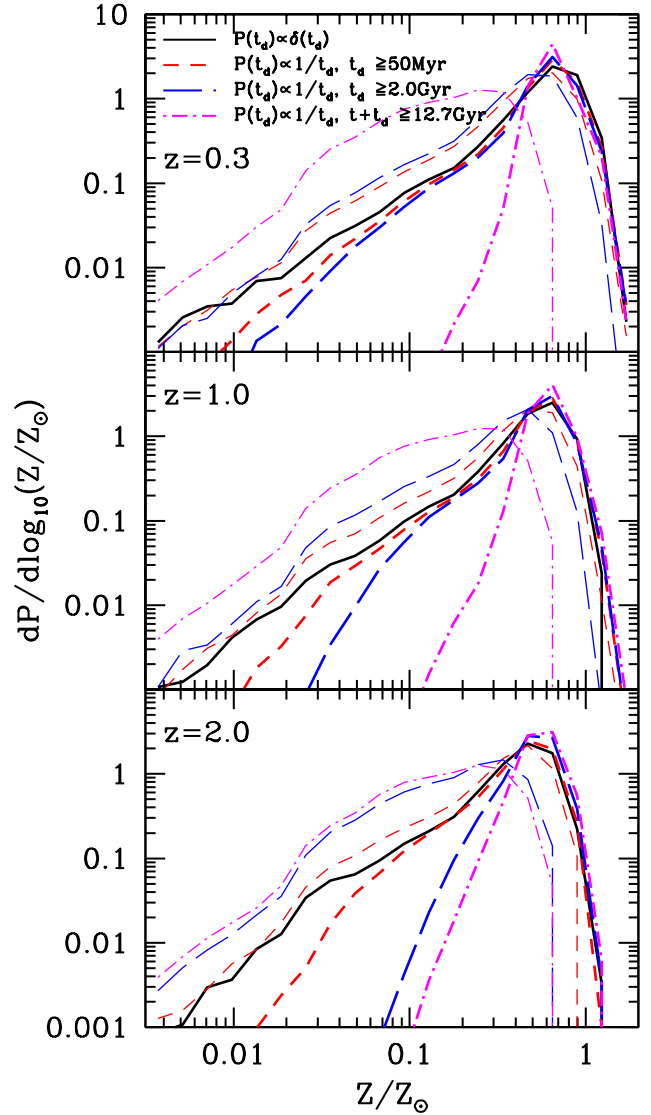}
\caption{Metallicity distribution of the host galaxies at the SBBH
formation time (thin lines) and the GW detection time or the SBBH
merger time (thick lines). Legends for the lines and models are
similar to Fig.~\ref{fig:f2}.
}
\label{fig:f7}
\end{center}
\end{figure}

Figure~\ref{fig:f7} shows the host galaxy metallicity distributions of
those SBBHs at their merger time (i.e., $z=0.3$, $1.0$, and $2.0$,
respectively; thick lines in panels from top to bottom) and at
formation time (thin lines), respectively. For those SBBH mergers
at $z=0.3$ produced from the reference model and the large time
delay model, the host galaxy metallicity distributions
at the SBBH formation time peak at $0.61 Z_\odot$
and $0.53 Z_\odot$, respectively. As expected, the larger the minimum
time delay, the poorer the metallicity of the host galaxies at the
SBBH formation time. The host galaxy metallicity distributions at
the SBBH merger time resulting from these two models shift to higher
metallicities due to the metal enrichment after the SBBH formation, and
the difference between the two distributions are relatively small.

If those SBBHs were formed at early time, e.g., $z\ga 6$ (the early
SBBH formation model), then they were mostly formed in metal poor
small galaxies with metallicity peaking at $0.26 Z_\odot$ and almost
half of them having $Z\la 0.12$ (thin dot-dashed magenta line). However,
their host galaxies at the SBBH merger time ($z=0.3$) have
metallicities around $Z\sim 0.57 Z_{\odot}$ with a $5$ to $95$
percentile range of $0.41$ to $ 0.84Z_{\odot}$. In this case, even if
the SBBH mergers were detected at high redshifts (e.g., $z=1$ or $2$),
their host galaxies are still quite metal rich ($Z\sim 0.39 - 0.73
Z_\odot$,$0.30 - 0.65 Z_\odot$ for $z=1$ or $2$ respectively), and
have metallicities on average only slightly lower than those detected
at lower redshift (e.g., $z=0.3$).

\begin{figure}
\begin{center}
\includegraphics[scale=0.45]{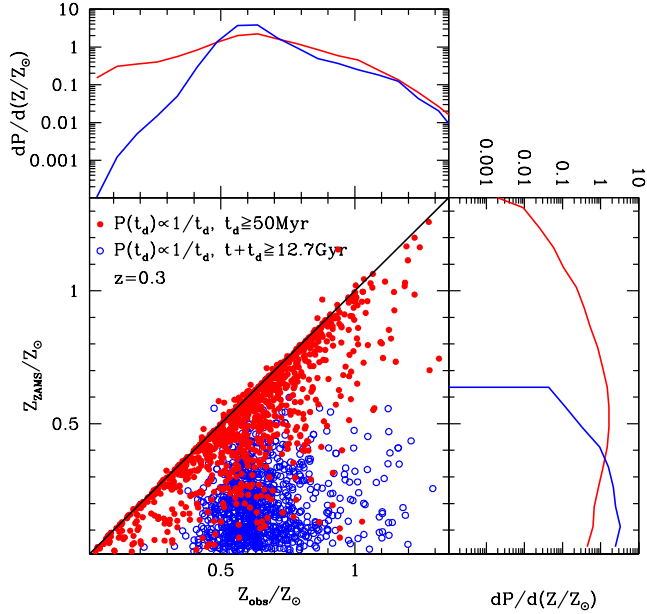}
\caption{Randomly selected mock SBBHs on the plane of the metallicity
of the host galaxy at the SBBH formation time versus the metallicity
of the host galaxy at SBBH merger time. Red filled circles and blue
open circles represent those mock SBBHs obtained form the early SBBH
formation model and the reference model, i.e.,  $P(t_{\rm d}) \propto
1/t_{\rm d}$ with $t_{\rm d} \ge 50\rm \,Myr$, and $t_{\rm d} \ge
12.7{\rm \,Gyr}- t(z)$, respectively.  }
\label{fig:f8}
\end{center}
\end{figure}

To further illustrate the difference between the metallicities of the
host galaxies at the SBBH formation time and that at the SBBH merger
time, in Figure~\ref{fig:f8} we plot those mock SBBH mergers at
$z=0.3$ obtained from the reference model (red filled circles) and the
early SBBH formation model (open blue circles) on the plane of the
host galaxy metallicities at the SBBH merger time ($Z_{\rm
obs}/Z_\odot$) versus those at the SBBH formation time ($Z_{\rm
ZAMS}/Z_\odot$). This figure clearly shows that the metallicities of
most SBBH host galaxies are enriched significantly after the formation
of SBBHs, and the longer the time delay of the SBBH mergers, the
more significant metal enrichment of the SBBH host galaxies.

According to Figures~\ref{fig:f7} and \ref{fig:f8}, the host galaxy
metallicity distribution of those SBBH mergers that were formed
at an early cosmic time in relatively metal poor galaxies is significantly less
extended at the low metallicity end than that for those SBBHs
formed at a later cosmic time in both metal rich and poor galaxies. This
seemingly surprise result is caused by that most of those
extremely metal poor host galaxies of SBBHs formed at early cosmic
time are significantly metal enriched by star formation later.

\begin{figure}
\includegraphics[scale=0.43]{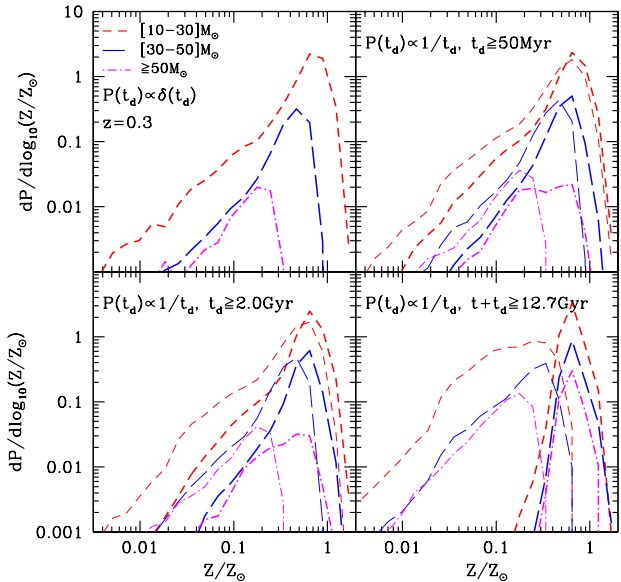}
\caption{Metallicity distribution of those SBBH mergers with different
mass ranges at both the SBBH detection time ($z=0.3$; thick lines) and
the SBBH formation time (thin lines). The red short dashed, blue
long-dashed, and magenta dot-dashed lines in each panel represent the
host galaxy metallicity distribution for mock SBBHs with total mass
$M_{\bullet\bullet}= 10-30\msun$, $30-50\msun$, and $\geq 50\msun$,
respectively. Panels from top to bottom, left to right, show the
results obtained from the prompt model, the reference model, the large
time delay model, and the early SBBH formation model, respectively.
}
\label{fig:f9}
\end{figure}

Figure~\ref{fig:f9} shows the metallicity distributions of the host
galaxies for SBBHs with different total mass ($M_{\bullet\bullet}$)
range (i.e., $10-30\msun$, $30-50\msun$, and $\ge 50\msun$) at both
the SBBH formation time and the SBBH merger time. For SBBH mergers
with $M_{\bullet\bullet}\ge 50\msun$,  they must be formed in galaxies
with metallicities $\la 0.2 - 0.3Z_\odot$ whenever their GW signals
were detected at low redshift (e.g., $z=0.3$) or high redshift (e.g.,
$z=2$) and whenever the time delays is longer or shorter, which is
primarily due to that the heavy SBBHs can only be formed from metal
poor binary stars. However, the metallicities of their host galaxies
at the SBBH merger time are all substantially metal richer with a $5$
to $95$ percentile range of $0.41$ to $0.84Z_\odot$ and have a
distribution skewed towarding high metallicities if those SBBHs were
formed at $z\ge 6$, or they have a $5$ to $95$ percentile range of
$0.07$ to $0.65Z_\odot$ (or $0.08$ to $0.65Z_\odot$) and a
distribution extending to low metallicities ($Z<0.1 Z_\odot$) if those
SBBHs were formed at a time $50$\,Myr (or $2$\,Gyr) before the
detection of their GW signals, or are also metal poor, i.e., $\la
0.2-0.3 Z_\odot$ if they merger shortly after their formation.

For light SBBHs, e.g., $M_{\bullet\bullet}=10-30M_\odot$, they can be
formed in galaxies with a large metallicity range (e.g., Z extends to
$1.1 Z_\odot$) whenever their GW signals are detected at low redshift
or high redshift and whenever the time delay is longer or shorter. At
the SBBH merger time, their host metallicities distribute over an even
larger range with broader extensions at both the high $Z$ and the low
$Z$ ends compared with that for $M_{\bullet\bullet} \geq 50\msun$ .
For SBBHs with intermediate mass [$M_{\bullet\bullet} \in
(30-50\msun)$], the distributions of the host metallicities at both
the SBBH formation time and the SBBH merger time are in between those
for the heavy SBBHs and for the light SBBHs (see Figure~\ref{fig:f9}).

\section{Conclusions and Discussions}
\label{sec:cons}

In this paper, we investigate the host galaxies of SBBH mergers, i.e.,
gravitational wave sources that may be detected by
aLIGO and advanced VIRGO (VIRGO), by implementing simple SBBH
formation recipes into cosmological galaxy formation model using the
Millennium-II simulation with a large box of side $137$\,Mpc, and we present
a complete and thorough analysis of the properties of the SBBH host galaxy.
Our main results are summarized as follows.

SBBH mergers with total mass $M_{\bullet\bullet}\ge 10\msun$ at low
redshift ($z\la 0.3$) occur preferentially in massive galaxies with stellar
mass $\ga 2\times10^{10} \msun$ if the delay time between the SBBH formation
and the merger is distributed in a broad range from less than Gyr to the Hubble
time, and they occur preferentially in even more massive galaxies ($\sim
8\times 10^{10}\msun$) if they were mostly formed at high redshift, e.g.,
$z>6$ and have a large delay time ($\ga 13$\,Gyr). For those heavy SBBH mergers ($M_{\bullet\bullet} \ge 50\msun$),
similar to GW\,150914, GW\,170104, and GW170814, their host stellar mass
distribution is probably bimodal, with a low mass peak of $\sim 10^9\msun$
and a high mass peak of $\sim 2 \times 10^{10}\msun$, if the delay time is
distributed in a broad range from less than Gyr to the Hubble time. The lower
peak is mainly contributed by those SBBHs formed recently ($z\la 0.5$), while
the higher mass peak is mainly contributed by those SBBHs formed at early
time ($z\ga 3.0$). If those
heavy SBBH mergers detected at low redshift (e.g., $z\la 0.3$) were
formed only at early time (e.g., $z>6$) or from Pop III stars, then
their host galaxies were originally small at the SBBH formation time
but became very massive  (mainly $\ga 3 \times 10^{10}\msun$) at the
GW detection time because of later assembly and growth, and the fraction of the host galaxies with mass $\la 10^9\msun$ is negligible ($\la 1\%$).

Some of these findings have been reported recently in \citet{Lamberts16}, \citet{Elbert17}, \citet{OShaughnessy17}, and \citet{Schneider17}.
For example, \citet{Lamberts16} found that GW\,150914-like heavy SBBHs are likely
to be found
in massive galaxies, but at a significant fraction of such mergers should be hosted in dwarf galaxies formed at low redshift ($z\sim 0.5$), roughly consistent with our results
for heavy SBBH mergers resulting from the reference model. Similar results may be also inferred from \citet{Elbert17} as
they stated that GW events would be preferentially localized in dwarf galaxies if the
merger timescale is short compared to the age of the universe and in massive galaxies
otherwise. \citet{OShaughnessy17} and \citet{Schneider17} also found that most
GW\,150914-like SBBH mergers should be hosted in star forming massive galaxies with
stellar mass $\ga 10^{10}\msun$ by implementing binary population synthesis models
into simulations of either several galaxies or a small box. In addition, \citet{Schneider17} also showed that fraction of GW\,150914 like events in low mass galaxies decreases,
but it is not zero. As the GW\,150914-like candidates produced in \citet{Schneider17} all
have very long delay time, their host galaxy distribution is more or less like the one shown by the magenta line in the  bottom right panel of Figure~\ref{fig:f6}.
Some of the quantitative differences of our results from those other studies may be
due to the different settings for the SBBH formation model. Further studies by utilizing large volume simulations,
such as Illustris \citep{2014MNRAS.444.1518V} and EAGLE
\citep{2015MNRAS.446..521S} with a large variety of mock galaxies, and implementation of binary population
synthesis would be necessary to accurately predict the properties
(including stellar mass and metallicity) of the host galaxies of SBBH
mergers and achieve better statistics.

We find that the host galaxies of GWs events detected at a high redshift
also have a broad mass distribution. The host galaxy mass distributions
at high redshifts obtained from different delay time models have similar
trends as those obtained at low redshift (e.g., $z=0.3$), and their peaks
shift toward lower masses and the fraction of host galaxies at the low (or
high) mass end increases (or decreases) with increasing GW detection redshift.

We find that the host galaxies of SBBH mergers with
$M_{\bullet\bullet} \ge 10\msun$ at low redshift (e.g., $z\la 0.3$)
are mostly spiral galaxies with stellar mass $\sim 5.3\times
10^{8}-6.0\times 10^{10} \msun$ and only $\sim$ ten percent are in
elliptical galaxies if the time period between the SBBH formation and
merger is distributed in a broad range from $\la$\,Gyr to the Hubble
time. However, about half of the hosts of those SBBH mergers detected
by GW at low redshift (e.g., $z\la 0.3$) should be elliptical galaxies
if those SBBHs were formed at early time or from Pop III stars (e.g.,
$z\ga 6$). The fraction of SBBH GW events that are hosted in
elliptical galaxies depends on when those SBBHs were formed and
merged, and it increases with the increasing of the time delay between
SBBH formation and merger.  We also find that most of the host
galaxies of SBBH mergers detected by GW at low redshift should be
central galaxies and only about $20\%$ of them are satellite galaxies,
which is almost independent of the settings of the delay time.

The formation of SBBHs depends on the metallicity of their progenitor
binary stars. Heavy SBBHs ($M_{\bullet\bullet} \geq 50\msun$) were
mostly formed in metal poor small galaxies with metallicities $\la
0.2-0.3Z_\odot$. If the time delay between SBBH formation and merger
covers a range from less than Gyr to the Hubble time, the host galaxy
metallicity distribution of SBBH mergers (with total mass
$M_{\bullet\bullet} \geq 10 \msun$) detected at redshift $z\la 0.3$
peaks around $0.6 Z_\odot$ and has a long skewed wing towarding
the low metallicity end ($Z< 0.2 Z_\odot$).  The host galaxy
metallicity distribution of heavy SBBHs (with total mass $\ga 50
M_\odot$, similar to GW\,150914, GW\,170104, and GW170814) at the
merger time at redshift $z\la 0.3$ is ranging from $Z<0.1Z_\odot$ to
$\sim 1.0Z_\odot$, substantially broader compared with that for less
heavy SBBHs, and the fraction of host galaxies of the heavy SBBHs with
$Z \la 0.2 Z_\odot$ is  much larger than that for less heavy SBBHs.
If SBBHs were formed at early time (e.g., $z>6$; or from Pop III
stars), their mergers detected at $z\la 0.3$ occur preferentially in
galaxies mostly with metallicities $\ga 0.2 Z_\odot$ and peaking
around $Z \sim 0.6 Z_\odot$, because of significant metal enrichment
of those host galaxies after the SBBH formation. It is interesting to
note here that the host galaxy metallicity distribution for the
mergers of those heavy SBBHs that were initially formed at an early
cosmic time (e.g., $z>6$) in metal poor galaxies, is significantly
less extended at the low metallicity end than that for those mergers
of SBBHs formed at a later cosmic time in both metal rich and poor
galaxies. This is seemingly a surprise result, as heavy SBBHs have to
be formed in more metal poor galaxies.

Searching for the EM counterparts of SBBH GW sources is important for
understanding the origin of those SBBHs and the application of them as
probes to cosmology. However, currently it is still difficult to find
these EM counterparts, if any, since the localization of these sources
by GW signals is poor. If the properties of the host galaxies of SBBH
mergers can be well understood as the efforts made in this study and
others \citep[e.g.,][]{Lamberts16, OShaughnessy17, Elbert17,
Schneider17}, the search for EM counterparts will be greatly narrowed
and the host galaxies of SBBHs may be statistically determined even
without EM counterpart information. Furthermore, if future
observations can discover the EM counterparts and host galaxies of
those SBBH mergers that will be detected by aLIGO, VIRGO, and other
future GW observatories, and thus may provide important data to
constrain the SBBH formation model and reveal the origin of those
SBBHs.

\section*{Acknowledgements}
This work is partly supported by the National Natural Science
Foundation of China under grant No. 11690024, 11373031 and 11390372, the
Strategic Priority Program of the Chinese Academy of Sciences (Grant
No. XDB 23040100), and the National Key Program for Science and
Technology Research and Development (Grant No. 2016YFA0400704).





%




\bsp	
\label{lastpage}
\end{document}